\documentclass[a4paper,11pt]{article}
\usepackage{pos}

\title{Main results of the TUS experiment on board the Lomonosov satellite}
 \ShortTitle{Results of the TUS experiment}

\author*[a]{Pavel Klimov}
\author[a]{Sergei Sharakin}
\author[a]{Mikhail Zotov}
\author[b]{Mario Bertaina}
\author[b]{Francesco~Fenu}
\forColl{Lomonosov-UHECR/TLE}
\affiliation[a]{Skobeltsyn Institute of Nuclear Physics of Lomonosov Moscow State University,\\
  Moscow, Russia}
\affiliation[b]{Università degli studi di Torino,
Via Pietro Giuria 1 10125, Torino, Italy}

\emailAdd{PavelKlimov@eas.sinp.msu.ru}

\abstract{The TUS detector was the first space-based mission aimed for
ultra-high-energy cosmic ray (UHECR) measurements. The detector was
designed to register the fluorescent signal of extensive air showers
(EAS) developing in the night atmosphere of Earth in the UV range of
300-400 nm. TUS was launched on board the Lomonosov satellite in April,
2016 and operated till December, 2017. Almost 90 thousand events were
recorded during the mission, among them lightning discharges, meteors,
transient luminous events, polar lights and anthropogenic signals. Some
puzzling bright UV flashes in a clear sky far from possible artificial
sources were also registered. Besides this, a number of EAS candidates
were found in the TUS database. The majority of candidates analysed so
far were recorded above populated areas near airports or similar
objects, and the energy of the signals corresponds to at least 1 ZeV if
they were generated by an UHECR, which does not allow one to consider
these events as UHECRs. We briefly present the main results of the TUS
experiment and discuss its importance for the development of the future
orbital missions.}

\FullConference{37$^{\rm{th}}$ International Cosmic Ray Conference (ICRC 2021)\\
		July 12th -- 23rd, 2021\\
		Online -- Berlin, Germany}

\begin{document}
\maketitle

\section{Introduction}

One of the main problems in ultra-high energy cosmic ray (UHECR) studies
is their extremely low flux at energies $\gtrsim5\times10^{19}$~eV,
above the so-called Greisen--Zatsepin--Kuzmin cut-off. This necessitates
running ground experiments occupying huge areas, which allows obtaining
sufficiently large exposures.  In early 1980's,
Benson and Linsley suggested to increase dramatically the exposure of
UHECR experiments by a totally different approach.  They suggested to
put a wide-field-of view fluorescence telescope into a low-Earth orbit
and employ it for registering fluorescence and Cherenkov emission of
extensive air showers born by UHECRs in the nocturnal
atmosphere~\cite{BL80}. TUS (Tracking Ultraviolet Setup), launched into
orbit on 28 April 2016, was the world's first telescope aimed to verify
the applicability of the idea by Benson and Linsley to UHECR studies.

\section{The TUS detector}

The design of TUS was developed in early
2000's~\cite{2001ICRC....2..831A}.
The telescope consisted of two main parts: a
mirror-concentrator and a photodetector in its focal plane.
The photodetector had 256 channels (``pixels'') with square windows of
$15\times15$~mm covering the field of view of $9^\circ \times 9^\circ$.
Pixel sensors were the photoelectron multiplier tubes (PMTs) R1463 of
Hamamatsu with a 13~mm diameter cathode. Near-UV (NUV) filters were
placed in front of PMTs in order to limit the wavelength range of
measurements to 300--400~nm.  Special light guides with square entrance
apertures (15~mm$\times$15~mm) and circular outputs were employed to
uniformly fill the detector’s field of view with PMT pixels. Sixteen
PMTs were combined into separate clusters (photodetector modules), and
each of the 16 clusters of the photodetector had its own digital data
processing system for the first-level trigger, based on a Xilinx FPGA as
well as a high-voltage power supply controlled by the FPGA to adjust the
PMT gains to the intensity of UV radiation.

Using PMTs as sensors and a Fresnel mirror with an area $\sim 2$~m$^2$
resulted in sensitive measurements with a high temporal resolution
(0.8~$\mu$s) and a spatial resolution about 5~km at sea level. A digital
oscilloscope implemented in the FPGA of the photodetector had four
operation modes with different time sampling windows: 0.8~$\mu$s,
25.6~$\mu$s, 0.4~ms and 6.6~ms.

An automatic gain control (AGC) system was an important part of the TUS
electronics. The PMT gains of were continuously (with a frequency of 20~Hz)
adjusted during the detector operation by tuning the high voltage so
that the average current in the anode circuit remained constant ($\sim
3~\mu$A). The PMT high voltage used to reach its maximum value
(1100~V) under conditions of the minimum background illumination.  The
voltage was lowered with an increasing illumination thus
decreasing the sensitivity of the PMTs.  This allowed an adjustment of
the detector to highly variable conditions of observation in nocturnal
segments of the orbit, see~\cite{SSR} for details.

A malfunction of the AGC system occurred during the first detector
turn-ons in the orbit. This was caused by the inoperability of the
voltage-lowering algorithm when the photodetector module was partially
exposed to a high-intensity light flux. In these conditions, a high
current took place in several PMTs bypassing the high voltage divider
and, thereby, lowering the signal in all other channels. As a result,
the detector spent some time on the day side of the orbit with maximum
gains before the AGC algorithm was corrected.  Due to this, two modules
and several channels in other modules of the photodetector became
completely inoperative, and sensitivities of other channels
decreased.
Together with the design solutions made in early 2000's, before the
discovery of the GZK cut-off, this resulted in an increase of the energy
threshold up to $\sim400$~EeV.

As another consequence, the pre-flight calibration became irrelevant and
it became necessary to develop and test a method for an in-flight
detector calibration, i.e., for determining PMT gains and sensitivity of
the data-acquisition channels basing on the data registered by the
detector.  The proposed method is based on the linear dependence of the
variance of the stationary signal on its baseline level (the mean
value): the slope of this dependence is proportional to the PMT gain.
The proportionality coefficient varies from one PMT to another and
depends on the position of the channel in the module; therefore, the
accuracy of the technique is low (the error is 30\% or larger),
see~\cite{Klimov2021} for details.


\section{Main results}

It became clear soon after the beginning of the TUS experiment that
all events in its data set can be roughly divided into four groups
basing on their phenomenology~\cite{JCAP2017}:
\begin{itemize}\parskip=-2pt

	\item events with stationary noise-like waveforms; these included
		events with a strongly non-uniform illumination of the focal
		surface with bright regions correlated with geographical positions
		of cities and objects like airports, power plants, offshore
		platforms etc.

	\item instant track-like flashes caused by charged particles
		hitting the UV filters of the photodetector;

	\item flashes produced by light coming outside of the FOV of the
		detector and scattered on its mirror; they were called ``slow
		flashes'' because of the long signal rise time in comparison
		with track-like flashes; and

	\item events with complex spatio-temporal dynamics; these included
		events with waveforms that could be expected from fluorescence
		originating from extensive air showers produced by
		extreme energy cosmic rays, ELVEs, as well as violent flashes
		of a yet unknown origin.

\end{itemize}

In what follows, we will briefly discuss the most remarkable groups
of events registered in two modes of operation.

\subsection{EAS-like events}

Some of the registered events have spatio-temporal pattern typical for
the EAS fluorescence: 

\begin{itemize}\parskip=-2pt

	\item a large number of hit pixels (channels whose signal significantly
		exceeds the background) that line up on the pixel map along some
		direction (a ``thick'' track);

	\item the total signal of the hit pixels with a subtracted background
		(the light curve) has a characteristic fast increase with the
		subsequent slightly slower decline (the full duration is about
		0.1~ms).

	\item noticeable time shifts of the signal peaks in the hit pixels
		indicating a motion of the image along the track.

\end{itemize}

The most interesting of these events, recorded on 3 October 2016 above
Minnesota, was analyzed in detail in paper~\cite{Minnesota} in an
assumption of its EAS origin.  In work~\cite{ICRC-2021_DustGrain}, the
event is considered in the context of a scenario of so-called
relativistic dust grains.

More than 120 EAS-like events were found in the data set registered by
TUS during its 1.5-year mission. Besides the signal registered above
Minnesota, five other events demonstrated a clear motion of the signal
similar to that expected from an EAS generated by an UHECR. However, all
of them were registered above the United States in a close vicinity of
airports,\footnote{Hereinafter TUS events are identified by indicating
the registration date, so the first event in Table~\ref{tab:TUSevents}
was triggered on 3 October 2016. To distinguish two events recorded
within one day, suffixes a and b are used.} see
Table~\ref{tab:TUSevents}.

\begin{table}[!ht]

	\caption{List of EAS-like events with a motion of the signal
	and their locations (geographical
	coordinates correspond to the center of the TUS FOV). The number of
	hit pixels is indicated in  the last column.}

	\begin{center}\small
		\begin{tabular}{|l|*{5}{c|}}
		\hline
		Event & Time (UTC) & Latitude & Longitude & Location  & \# hit pixels\\
		\hline
		TUS161003 & 05:48:59 & 44.08$^\circ$N & 92.71$^\circ$W & Minnesota & 10  \\
		\hline
		TUS161031 & 10:25:18 & 61.30$^{\circ}$N & 155.69$^{\circ}$W & Alaska  & 8\\
		\hline
		TUS170915 & 06:30:18 & 40.31$^{\circ}$N & 107.07$^{\circ}$W & Colorado & 12 \\
		\hline
		TUS171010 & 04:26:04 & 34.83$^{\circ}$N & 77.39$^{\circ}$W & North Carolina & 15 \\
		\hline
		TUS171029a & 06:39:09 & 35.27$^{\circ}$N & 110.78$^{\circ}$W & Arizona  & 8\\
		\hline
		TUS171029b & 11:13:26 & 65.90$^{\circ}$N & 168.07$^{\circ}$W & Alaska & 9 \\
		\hline
		\end{tabular}

	\end{center}
	\label{tab:TUSevents}

\end{table}

\begin{figure}
    \centering
    \includegraphics[width=0.45\textwidth]{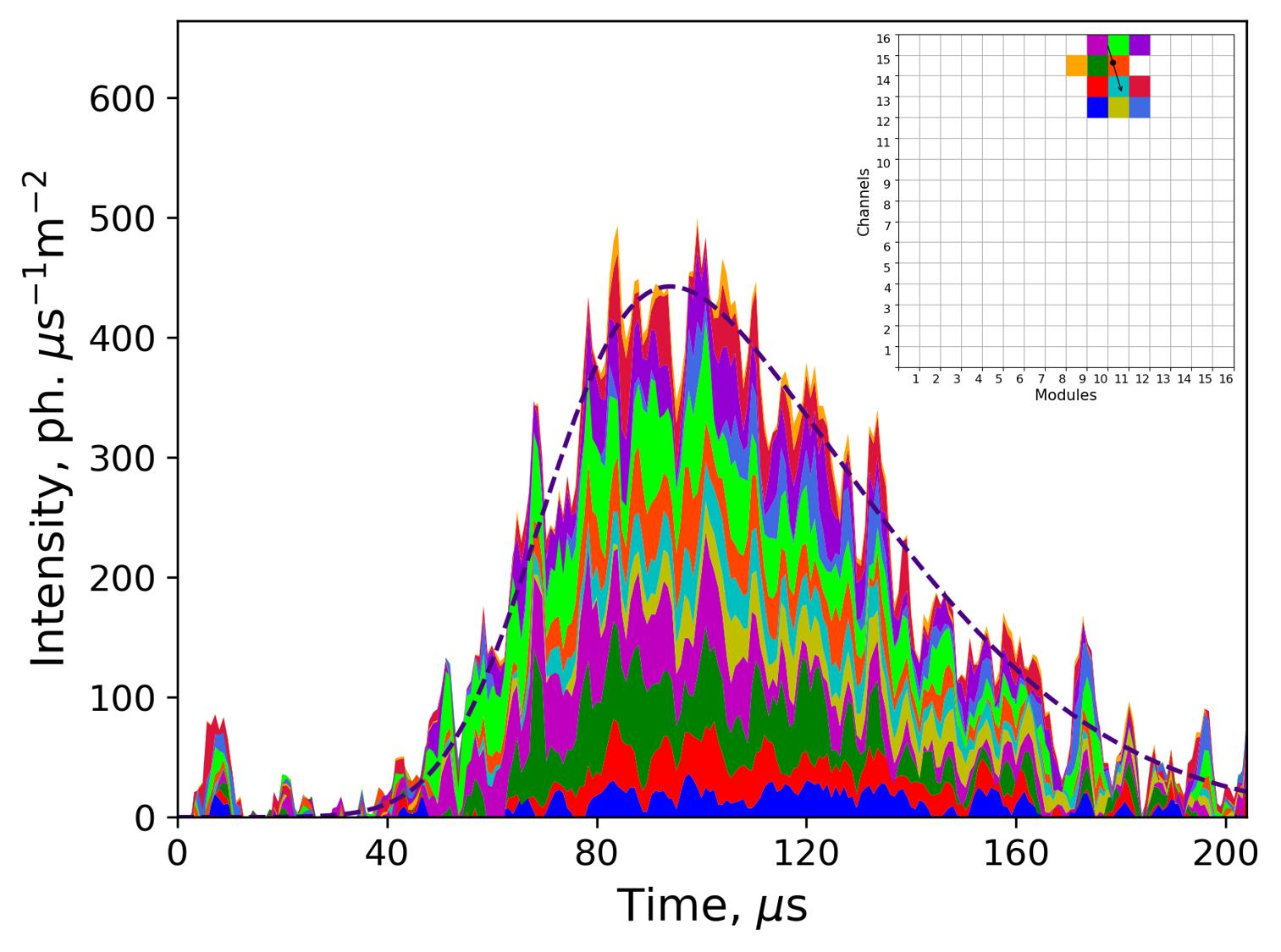}
    \includegraphics[width=0.5\textwidth]{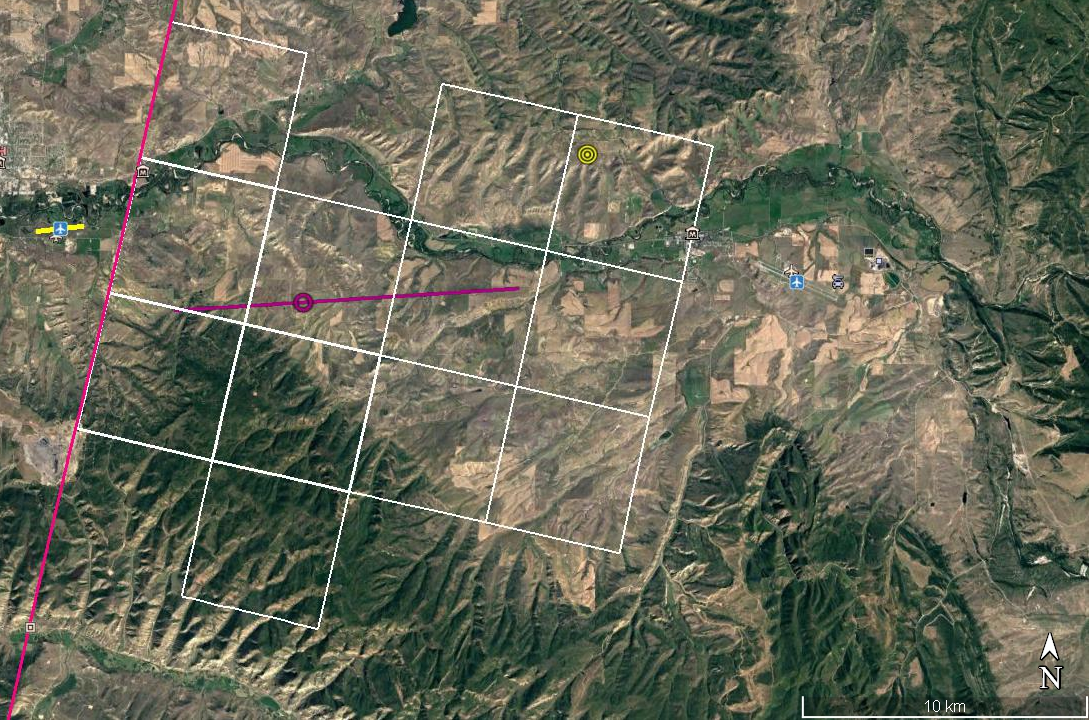}

	 \caption{EAS-like event TUS170915. Left: the light curve and the
	 pixel map.  Right: the reconstructed track direction (the violet
	 line) and the runway of the Craig Moffat County airport (the yellow
	 segment).}

    \label{fig:sample_LC}
\end{figure}

A typical light curve can be seen in the left panel of
Fig.~\ref{fig:sample_LC} (colors represents different hit pixel
signals).  The duration of the EAS-like events at half-amplitude is in
the range of 50--70~$\mu$s, and the track length when projected onto the
earth's surface varies from~5 to 12~km. It should be noted that the
motion of the source has a pronounced relativistic character: the
reconstructed angular velocity of movement of the source is from 100 to
200~rad/s, which corresponds to the component of the linear velocity
across the line of sight from $0.1c$ to $0.2c$ ($c$ is the speed of
light). The reconstructed values of the zenith angles, assuming a source
velocity equal to~$c$, lie in the range from 11$^\circ$ to 38$^\circ$.


The origin of these EAS-like events is still unclear. It is difficult to
interpret them as an EAS fluorescence due to their significant signal
amplitudes: they vary in the range from 120 to 450~photons per m$^2$ per
$\mu$s, which corresponds to a primary energy of a proton the order of
1~ZeV and higher. The probability of registering such an UHECR is of the
order of $10^{-3}$--$10^{-5}$, see~\cite{Minnesota}. Therefore, a
special attention was paid to the search of possible artificial sources.
In particular, a strong correlation was found between the reconstructed
tracks and the location of nearest airports for 5 out of 6 events. The
right panel of Fig.~\ref{fig:sample_LC} shows a Google map with a part
of the field of view of the TUS170915 event, the reconstructed track
indicated by a violet line with a circle (``maximum point''), and the
Craig Moffat County airport runaway indicated by a yellow segment. White
squares show boundaries of the fields of view of the hit pixels.


\subsection{Polar lights}

Due to the telescope construction providing the high sensitivity and the
spacecraft polar orbit, the TUS detector observed NUV pulsations of the
atmospheric glow at high latitudes: the threshold energy of the UV
emitted from the atmosphere measurements of the TUS detector is less
than the average intensity of auroral luminosity.
The observed signals have a very diverse structure with characteristic
frequencies of the order of 1--10~Hz.  The luminescence regions are
localized spatially with a characteristic size about 10~km. Several
different pulsation regions with different temporal structures
(waveforms) were observed simultaneously in the FOV of the telescope. 

\begin{figure}
	\centering
	\includegraphics[width=0.85\textwidth]{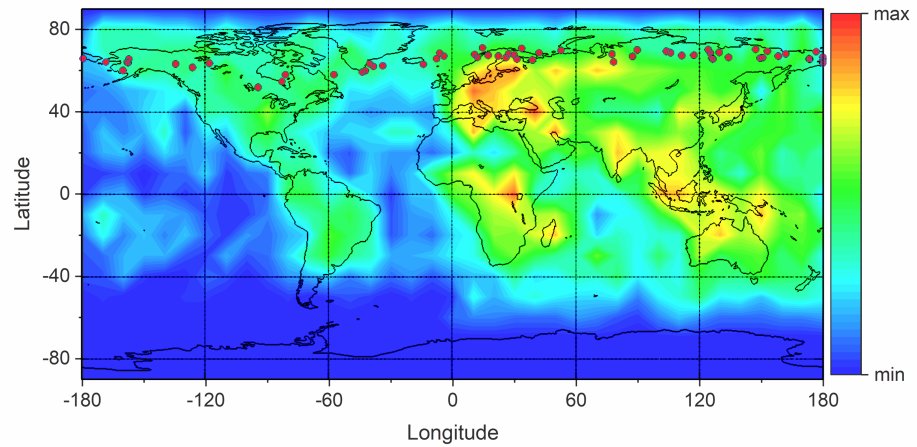}

	\caption{Map of selected events with NUV pulsations (red dots). The
	colour diagram shows the distribution of all events registered by the
	detector.}

	\label{geo}
\end{figure}

The geographic distribution and geomagnetic conditions analysis
indicates that the events are measured at the equatorial border of the
auroral zone, see Fig.~\ref{geo}. Their location does not depend on
geomagnetic activity level which is typical for the other auroral
events. The maximum portion of the pulsations is recorded in the
L-shells range from~4 to~6. Although, the event's occurrence frequency
correlates with geomagnetic activity. 

\begin{figure}[!ht]
	\centerline{%
		\includegraphics[width=0.4\textwidth]{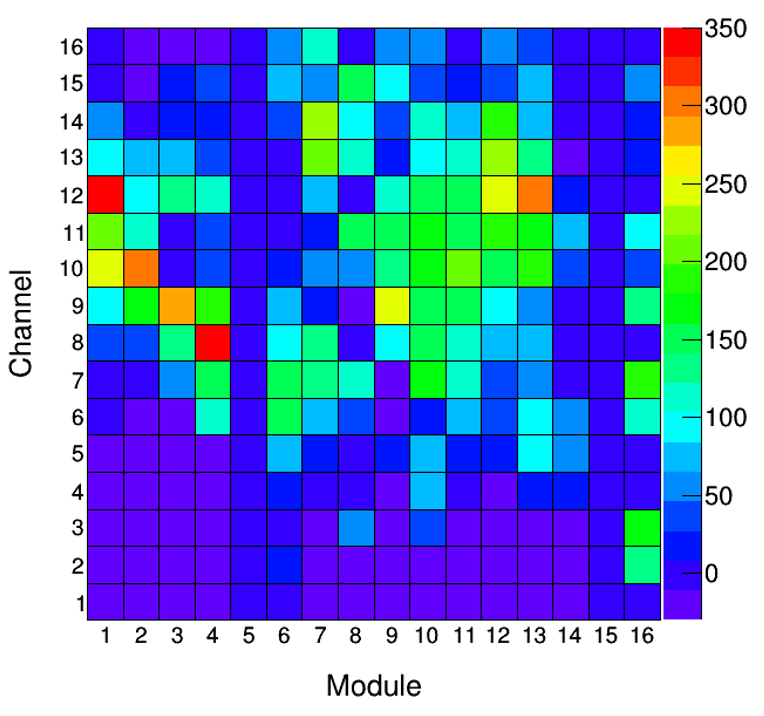}
		\includegraphics[width=0.52\textwidth]{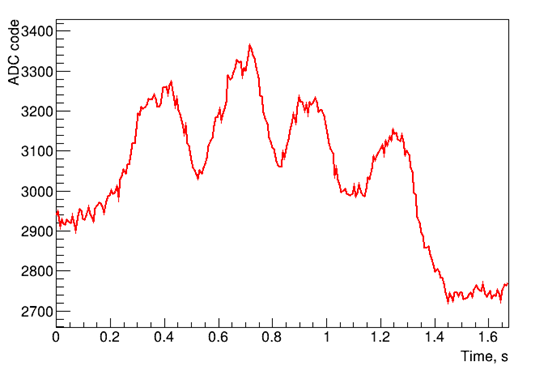}
		}

	\caption{An event with NUV pulsations measured by the TUS
	detector on 10 November 2017 at 13:31~UTC.}

	\label{Aurora_NUV}
\end{figure}

In terms of the event spatio-temporal structure, these signals are
similar to flickering auroras observed earlier~\citep{Sakanoi2005} and
internal modulations of pulsating auroras related to a high-energy part
of precipitating electrons caused by LBC waves. An example of one event
(one of the waveforms and a pixel map) is shown in Fig.~\ref{Aurora_NUV}
However, the nature and the occurrence mechanism of such signals is not
obvious yet. A further research based on high-sensitive orbital detector
observations is required to obtain detailed characteristics of these
events. A comparison with optical data of ground-based geophysical
observatories and satellite data on charged particles fluxes will help
to clarify the nature of the phenomenon. The TUS detector had a 2~m$^2$
mirror which provided a very high sensitivity of the detector to measure
faint emissions which can not be registered by ground-based all-sky
cameras. High sensitivity and temporal resolution allows such a detector
to measure rare events with faster pulsations (for example, such as
those presented in~\cite{Kataoka2012}) if a sufficient exposure is
achieved.

\subsection{Elves}

Elves (``ELVE'' stands for Emission of Light and Very Low Frequency
perturbation from an Electromagnetic Pulse) are the type of transient
luminous events (TLEs) that represent expanding luminous rings in the
ionosphere at the height of 80--90~km. The duration of an elve is less
than 1~ms and they can expand over 300~km laterally. It is believed that
they are the result of ionospheric electrons heating by the upward
electromagnetic impulse radiated by the lightning discharge
current~\cite{Inan1991}.  According to the ISUAL global experimental
data~\cite{isual2008}, elves are the most common type of TLEs: around
50\% of all TLEs were found to be elves.

\begin{figure}[!ht]
   \centering
   \includegraphics[height=.35\textwidth]{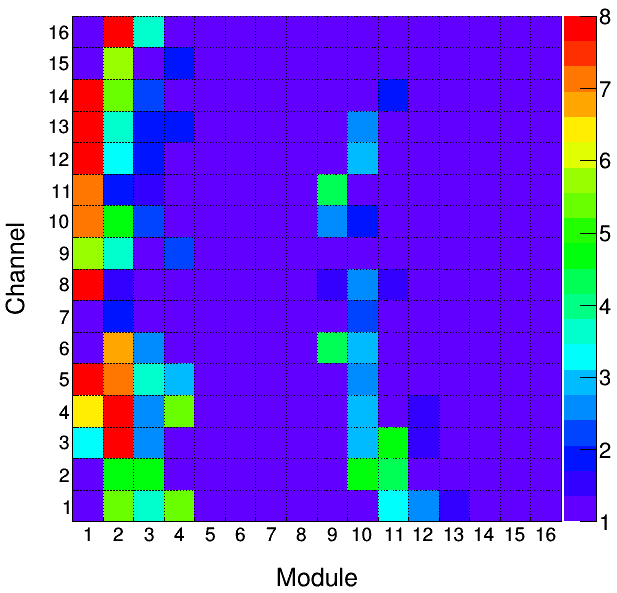}\quad
   \includegraphics[height=.35\textwidth]{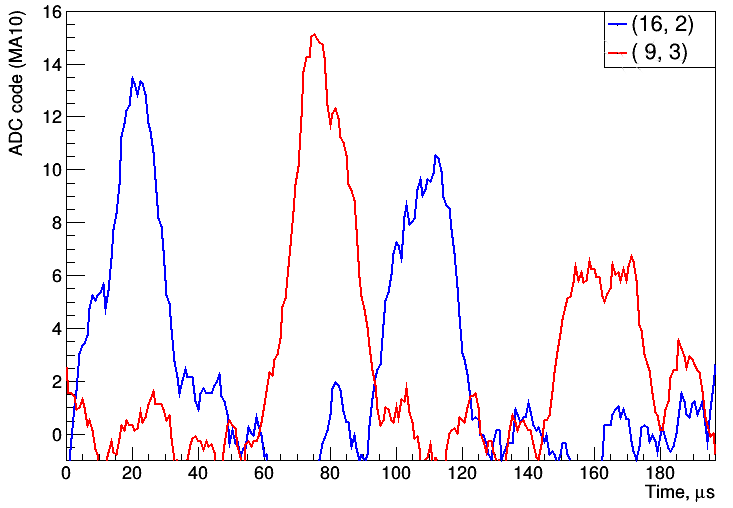}

	\caption{A double elve measured by TUS on 10 April 2017.
	Left: a pixel map with two bright arcs. Right: waveforms of two
	channels (blue and red lines), each comprising two peaks with a time
	delay of $\sim85~\mu$s.  }

	\label{elve}
\end{figure}

Usually ordinary (single) elves are caused by a cloud-to-ground
lightning of any polarity.	 The TUS detector measured at least twenty
six events that can be classified as elves.  Several of them
have a more complicated space-time pattern: two or more rings were
moving with a high speed across the field of view. An example of such a
``double elve'' is presented in Fig.~\ref{elve}.  Two separate rings are
clearly seen on the pixel map (the signal was integrated over
6.4~$\mu$s). These rings correspond to two peaks in the waveforms.
Signals of two channels are given for comparison in the right panel of
the figure. The first ring is brighter.  It arises as a result of the
action of a direct electromagnetic impulse from a lightning on the
ionosphere. The second ring is caused by a pulse reflected from the
ground.

\section{Discussion and Conclusions}

The TUS detector was the first space-based mission aimed for
ultra-high-energy cosmic ray measurements. Its mission lasted from 19
May 2016 till 30 November 2017, and the total exposure reached
$\sim$1550~km$^2$~sr~yr~\cite{Fenu2021}. A number of EAS-like events
were measured that have a typical light curve and demonstrate a movement
in the FOV at a relativistic velocity. The energy evaluation for these
events (for example, TUS161003 event) provides a value of one order of
magnitude higher than could be expected from UHECRs. This can be
interpreted in several ways. Most probable is the anthropogenic origin
of the events.  However, an astrophysical hypothesis associated with
relativistic dust grains is also considered.

The TUS detector used to register various UV phenomena which constitute
the background for UHECR measurements. Among them are anthropogenic
lights, thunderstorm activity and lightning discharges, upper atmosphere
transient luminous events, polar lights etc. The analysis of the TUS
data is still ongoing. One of the recent studies includes an application
of neural networks to classification of its data~\cite{universe}.

The orbital fluorescent technique proved a possibility to measure and
recognize a relativistic motion in the UV range in the atmosphere, to
reconstruct the direction and energy of the event. On the other hand,
the experience of the TUS mission revealed difficulties of a space-based
experiment that needs an accurate monitoring of the rapidly changing
background illumination and a high-quality control of the sensitivity of
the equipment.

The TUS detector demonstrated a multi-functionality of an orbital
fluorescent observatory and its usefulness for various astrophysical and
geophysical studies.  It provided an invaluable experience for the
implementation of the future orbital missions like
K-EUSO~\cite{K-EUSO2017} and POEMMA~\cite{POEMMA2021}.

\section*{Acknowledgments}

This work was supported by Space State Corporation ROSCOSMOS, and by
Lomonosov Moscow State University in frame of Interdisciplinary
Scientific and Educational School of Moscow University ``Fundamental and
Applied Space Research.'' The Italian group acknowledges financial
contribution from the agreement ASI-INAF n.2017-14-H.O.

\bibliography{references}

\newpage
\noindent
{\Large\bf The Lomonosov-UHECR/TLE Collaboration\\}

G.K. Garipov$^a$,
\fbox{B.A.~Khrenov$^a$},
P.A.~Klimov$^a$,
\fbox{M.I.~Panasyuk$^{a,b}$},
V.L.~Petrov$^a$,
S.A.~Sharakin$^a$,
A.V.~Shirokov$^a$,
I.V.~Yashin$^a$,
M.Yu.~Zotov$^a$,
A.A.~Grinyuk$^c$,
V.M.~Grebenyuk$^c$,
M.V.~Lavrova$^c$,
L.G.~Tkachev$^{c,d}$,
A.V.~Tkachenko$^c$,
O.A.~Saprykin$^e$,
\fbox{A.A.~Botvinko$^e$},
A.N.~Senkovsky$^e$,
A.E.~Puchkov$^e$,
M.~Bertaina$^f$,
F.~Fenu$^f$

{ \footnotesize
\noindent
$^a$Lomonosov Moscow State University, Skobeltsyn Institute
of Nuclear Physics, GSP-1, Leninskie Gory, Moscow, 119991, Russia\\
$^b$Physics Department, Lomonosov Moscow State University,
Leninskie Gory, Moscow, 119991, Russia\\
$^c$Joint Institute for Nuclear Research,
Joliot-Curie, 6, Dubna, 141980, Moscow region, Russia\\
$^d$Dubna State University,
University str., 19, Bld.1, Dubna, Moscow region, Russia\\
$^e$Space Regatta Consortium,
ul. Lenina, 4a, Korolev, 141070, Moscow region, Russia\\
$^f$Universit\`a degli studi di Torino,
Via Pietro Giuria 1, 10125 Turin, Italy\\
}
\end{document}